\documentclass[cameraready]{Interspeech}

\title{FastWave: Optimized Diffusion Model for Audio Super-Resolution}

\author[affiliation={1,2}, orcid=0009-0005-1670-1906, equalcontribution]{Nikita}{Kuznetsov}
\author[affiliation={1}, orcid=0000-0002-0925-3130, equalcontribution, correspondingauthor]{Maksim}{Kaledin}

\address{
    $^1$ HSE University, 109028 11 Pokrovsky bulvar, Moscow, Russia\\ 
    $2$ VK LLC, 125167 39 Leningradsky prospekt, bld 79, Moscow, Russia}

\email{nvkuznetsov\_4@edu.hse.ru, mkaledin@hse.ru}


\usepackage{comment}

\usepackage{amsmath}
\usepackage{amssymb}
\usepackage{mathtools}    
\usepackage{graphicx}
\usepackage{hyperref}

\begin{document}

\maketitle

\begin{abstract}
    Audio Super-Resolution is a set of techniques aimed at high-quality estimation of the given signal as if it would be sampled with higher sample rate. Among suggested methods there are diffusion and flow models (which are considered slower), generative adversarial networks (which are considered faster), however both approaches are currently presented by high-parametric networks, requiring high computational costs both for training and inference. We propose a solution to both these problems by re-considering the recent advances in the training of diffusion models and applying them to super-resolution from any to 48 kHz sample rate.  Our model called FastWave has around 50 GFLOPs of computational complexity and 1.3 M parameters and can be trained with less resources and significantly faster than the majority of recently proposed diffusion- and flow-based solutions. FastWave also has comparable performance to state-of-the-art models. We provide our implementation on GitHub\footnote{\texttt{https://github.com/Nikait/FastWave}}.
\end{abstract}

\section{Introduction}
Audio super-resolution problem consists in estimating missing high-frequency components of an audio signal. Roughly speaking, one would like to resample low-resolution audio (e.g. 8 kHz) recorded in limited setting into a high-resolution one (e.g. 48 kHz) to improve perceptual quality of the signal. Despite the fact that there are computationally very cheap interpolation approaches, they still cannot deliver sufficient perceptual quality in high-frequency band (above the original signal's Nyquist frequency) \cite{flowhigh2025}. Deep Learning (DL) approaches entered the field just in recent years, resulting in well-developed solutions. The majority of suggested approaches focus on the quality of the signal and rarely address the complexity, especially this is true for diffusion models still suffering from slow inference. This becomes critical in low-resource setting of consumer devices, where the ability to do edge computing (i.e. on device) is very valuable \cite{Li2022-mv}.\\

Our main contributions are the following.
\begin{enumerate}
    \item We develop one of  the smallest available diffusion model for audio super-resolution in the literature by optimizing NU-Wave 2 model using new-generation convolution blocks. Our final model has only 1.3 M parameters, showing 30\% decrease in parametric complexity.
    \item We optimized the training methodology of NU-Wave 2 via changing the paradigm to denoising and introducing various methodologies from EDM \cite{edm2022}. This allowed us to reach the same or better results in more constrained setting and less training iterations.
    \item Our model is capable of transforming audio from any sample rate to 48 kHz. The results are demonstrated on benchmark super-resolution problem on VCTK dataset and our methods are compared to the modern state-of-the-art solutions.
\end{enumerate}

\section{Related Work}
DL approaches achieved a significant success in audio super-resolution, showing impressive performance in improving the perceptual quality of the speech and low-resolution media content. Discriminative approaches \cite{AECNN2021} typically involve 10 M parameters and several times more to address the problem of super-resolution with varying input \cite{liuNVSR2022}. Switching the paradigm to generative adversarial training (GAN) introduced new possibilities of reaching good performance \cite{frepainter2024} and in the same time demonstrated that the computational complexity of the model can be significantly decreased \cite{hifipp}. GAN approaches addressed not only the quality \cite{apbwe}, but also the computational complexity and inference speed \cite{baenet}.

Apart from GANs diffusion models also entered the field just recently with NU-Wave \cite{nuwave}, having moderate complexity (around 3 M parameters) and close-to-SOTA results. The solution was developed into NU-Wave 2 \cite{nuwave22022} with changed architecture, any-to-48 kHz flexible-input regime and two times smaller model. Despite these findings, large part of diffusion-based \cite{audiosr,udmplus} solutions were focused mainly on achieving better reconstruction error or perceptual evaluation metrics. Flow-based models mainly follow the same path in terms of complexity, but they got a considerable advantage due to its one-step nature \cite{flowhigh2025}. As was demonstrated in \cite{nuwave22022}, sufficient computational resource is required to train the model to its peak results. Since NU-Wave 2, the question of constructing a smaller (in terms of parameters), faster (in terms of number of function evaluation (NFE) and computational complexity) diffusion-based model and with less training efforts remains open. On the other hand, in the field of image processing there was developed the new methodology for training diffusion-based models, called EDM \cite{edm2022,edm22024} which promised new optimized training methodology aimed at the reduction of training iterations. Our paper addresses all these challenges: low-parametric model, NFE reduction and optimized training time using the EDM methodologies.

\section{Methodology}
Let $X \in \mathbb{R}^{T}$ be a monaural audio signal. The goal of super-resolution is to reconstruct $X$ from its low-resolution counterpart obtained via $p$-times downsampling:
\begin{equation}
Y = \mathrm{Down}(X, p) \in \mathbb{R}^{\lceil T / p \rceil}.
\end{equation}
We estimate the high-resolution signal by learning a parametric mapping given by a diffusion model pipeline $f_\theta$:
\begin{equation}
\hat{X} = f_\theta(Y) \approx X.
\end{equation}

We build upon NU-Wave 2 \cite{nuwave22022}, which is among the most parameter-efficient diffusion models for audio super-resolution. We consider three successive model variants:
\begin{enumerate}
    \item NU-Wave 2 (baseline) — the original model;
    \item NU-Wave 2 + EDM — the baseline, trained and sampled using the EDM framework;
    \item FastWave — the final model, NU-Wave 2 + EDM diffusion modeling~\cite{edm2022} with architectural improvements given by ConvNeXtV2~\cite{Woo2023ConvNeXtV2}.
\end{enumerate}

\subsection{FastWave}
\subsubsection{Overview}
\begin{figure}
    \centering
    \includegraphics[width=0.95\linewidth]{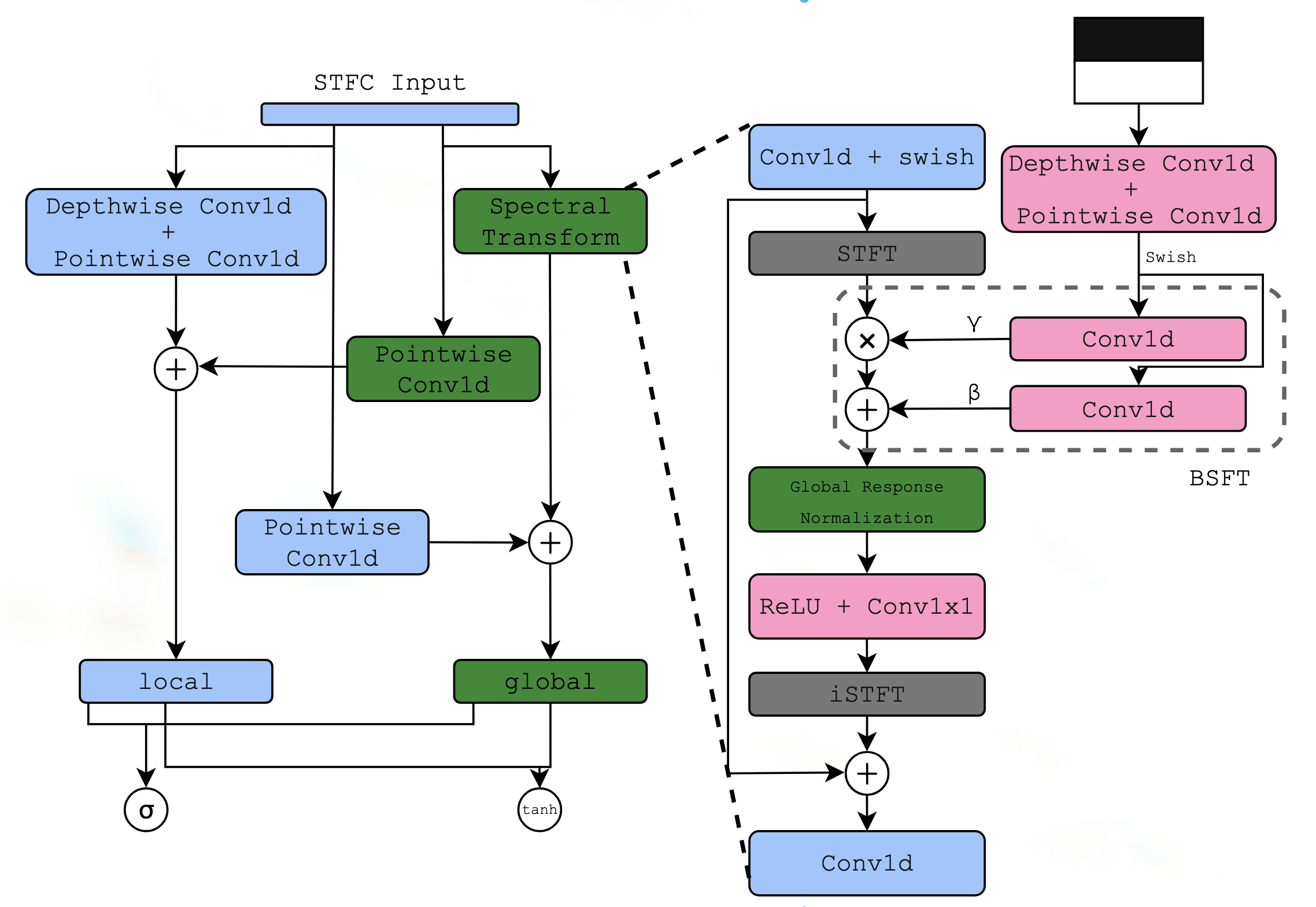}
    \caption{Architecture of FastWave with proposed architectural improvements.}
    \label{fig:fastwave_arch}
\end{figure}

In comparison to NU-Wave 2, FastWave replaces the original diffusion formulation with a denoising structure with $\sigma$-parameterization, as in EDM. The architecture of the main STFC and BSFT blocks has also been modified (see Figure ~\ref{fig:fastwave_arch}). The detailed description is provided in \hyperref[subsec:arch_modifications]{Subsection~\ref*{subsec:arch_modifications}}.

\subsubsection{Diffusion Parameterization}
Instead of predicting noise $\epsilon$ as in NU-Wave 2, FastWave is trained as a denoiser
\begin{equation}
D_\theta(x + n; \sigma) \approx x, \quad n \sim \mathcal{N}(0, \sigma^2 I),
\end{equation}
where $\sigma$ directly controls the noise level. The corresponding score function is
\begin{equation}
\nabla_x \log p(x;\sigma) = \frac{D_\theta(x;\sigma) - x}{\sigma^2}.
\end{equation}

\subsubsection{Network Preconditioning}
Following EDM, we apply explicit input--output preconditioning:
\begin{align}
x_{\text{in}} &= c_{\text{in}}(\sigma)\, x, \\
c_{\text{in}}(\sigma) &= \frac{1}{\sqrt{\sigma^2 + \sigma_{\text{data}}^2}}, \\
D_\theta(x;\sigma) &= c_{\text{skip}}(\sigma)\, x + c_{\text{out}}(\sigma)\, F_\theta(x_{\text{in}}, \sigma),
\end{align}
with
\begin{equation}
c_{\text{skip}}(\sigma) = \frac{\sigma_{\text{data}}^2}{\sigma^2 + \sigma_{\text{data}}^2}, \quad
c_{\text{out}}(\sigma) = \frac{\sigma\,\sigma_{\text{data}}}{\sqrt{\sigma^2 + \sigma_{\text{data}}^2}}.
\end{equation}
The parameter $\sigma_{\text{data}}$ is estimated directly from the training dataset by computing the empirical standard deviation of the data.

\subsubsection{Training Objective}
FastWave is trained using a weighted L2 denoising loss:
\begin{equation}
\mathcal{L} =
\mathbb{E}_{x,n,\sigma}
\left[
\lambda(\sigma)\,
\lVert D_\theta(x+n;\sigma) - x \rVert_2^2
\right],
\end{equation}
where
\begin{equation}
\lambda(\sigma) = \frac{\sigma^2 + \sigma_{\text{data}}^2}{(\sigma\,\sigma_{\text{data}})^2}.
\end{equation}
The noise level is sampled from a log-normal distribution:
\begin{equation}
\ln \sigma \sim \mathcal{N}(P_{\text{mean}}, P_{\text{std}}^2).
\end{equation}

\noindent\textbf{Choice of $P_{\text{mean}}$ and $P_{\text{std}}$.} The hyperparameters $P_{\text{mean}}$ and $P_{\text{std}}$ are chosen in a data-driven manner by approximating the mean and variance of $\text{ln} \sigma$ over the entire training dataset. This allows the log-normal noise distribution to concentrate sampling on intermediate noise levels, where the denoising loss is most informative, as in \cite{edm2022}.

\subsubsection{Sampling}
During inference, FastWave follows the probability flow ODE formulation and employs a first-order Euler solver. In contrast to NU-Wave~2, which relies on a fixed log-SNR schedule, we adopt the continuous noise schedule proposed in EDM:
\begin{equation}
\sigma_i =
\left(
\sigma_{\max}^{1/\rho}
+
\frac{i}{N-1}
\left(
\sigma_{\min}^{1/\rho} -
\sigma_{\max}^{1/\rho}
\right)
\right)^\rho,
\quad i = 0, \dots, N-1,
\end{equation}
where $\sigma_{\max}$ and $\sigma_{\min}$ denote the maximum and minimum noise levels, and $\rho$ controls the concentration of steps at low noise values.

Starting from $x_N \sim \mathcal{N}(0, \sigma_{\max}^2 I)$, we iteratively update the sample using the Euler discretization of the probability flow ODE:
\begin{equation}
x_{i-1} = x_i + (\sigma_{i-1} - \sigma_i) \frac{D_\theta(x_i;\sigma_i) - x_i}{\sigma_i}.
\end{equation}

\subsubsection{Architectural Modifications}
\label{subsec:arch_modifications}
To further reduce model complexity of NU-Wave 2 while preserving expressive capacity, we adopt several architectural changes inspired by ConvNeXtV2~\cite{Woo2023ConvNeXtV2}.

\noindent\textbf{Depthwise separable convolutions.}
In the original architecture, most local processing blocks relied on standard convolutions of the form
\begin{equation}
\texttt{Conv1d}(C_{\text{in}}, C_{\text{out}}, k),
\end{equation}
which scale quadratically with the number of channels. Following ConvNeXt, we replace them with depthwise separable convolutions, decomposed into
\begin{equation}
\texttt{DWConv1d} (C_{\text{in}}, k) \rightarrow \texttt{PWConv1d} (C_{\text{in}}, C_{\text{out}}, 1).
\end{equation}
This change significantly reduces the number of parameters and FLOPs, especially for large channel counts $N$, while retaining a comparable receptive field. Specifically, depthwise convolutions are introduced in the local branch of the FFC module and the BSFT shared MLP block.

\noindent\textbf{Global Response Normalization.}
Following ConvNeXtV2, we introduce Global Response Normalization after depthwise or expanded-channel transformations. GRN explicitly normalizes responses across channels and improves cross-channel interaction, which is especially important when depthwise convolutions limit channel mixing. GRN layers are inserted after the shared BSFT MLP and before the output projection in each residual block.

\section{Metrics}
We evaluate the proposed method using both reconstruction and computational complexity metrics.

\subsection{Reconstruction quality metrics}
We use Signal-to-Noise Ratio (SNR) and Log-Spectral Distance (LSD) defined as following:
\begin{equation}
\mathrm{SNR} = 10 \log_{10} \left( \frac{\sum_{n} x(n)^2}{\sum_{n} \left( x(n) - \hat{x}(n) \right)^2} \right),
\end{equation}
\begin{equation}
\mathrm{LSD} = \frac{1}{K} \sum_{k=1}^{K}
\sqrt{
\frac{1}{F} \sum_{f=1}^{F}
\left(
20 \log_{10} \frac{\left| X_k(f) \right|}{\left| \hat{X}_k(f) \right|}
\right)^2
},
\end{equation}
The values $x(n)$ and $\hat{x}(n)$ denote the original and reconstructed signal with $X_k(f)$ and $\hat{X}_k(f)$ being their magnitude spectra. Number $K$ represents the number of time frames, $F$ is the number of frequency bins. We additionally report LSD-LF and LSD-HF \cite{nuwave22022}, computed over the low-frequency and high-frequency bands, respectively. In our implementation, the frequency cutoff between low- and high-frequency regions is defined according to the input sampling rate. Specifically, we compute the short-time Fourier transform (STFT) using a 2048-point FFT, which results in 1025 frequency bins. The cutoff index is computed as
\begin{equation}
f_c = \left\lfloor 1025 \cdot \frac{f_{\mathrm{in}}}{48000} \right\rfloor,
\end{equation}
where $f_{\mathrm{in}}$ is the input sampling rate and $48000~\mathrm{Hz}$ is the full-band sampling rate. LSD-LF is computed by restricting the summation over frequency bins to $[0, f_c)$, while the LSD-HF is computed over $[f_c, 1025)$.

\subsection{Complexity metrics}
To evaluate computational efficiency, we report the Real-Time Factor (RTF), which characterizes the inference speed of the model relative to the duration of the processed audio. In addition, we report the number of GFLOPs per second of audio and the number of trainable parameters.

\section{Experimental Setup}
We conducted our experiments on the VCTK dataset \cite{valentinibotinhao16_interspeech}.
This dataset is a sample of 110 speakers from the VoiceBank corpus, primarily recordings of real speech spoken by native English speakers with a variety of accents.
The original dataset sampling rate is 48 kHz, making the entire dataset approximately 44 hours of audio.
We performed the usual speaker-specific holdout, using 100 speakers for training and 8 speakers for the test set. 
All models in the tables were measured on this portion of the dataset for consistency. We compared all models on the benchmark of upsampling into 48 kHz from 8, 12, 16 and 24 kHz.

To validate the effect of new methodologies, we compared the NU-Wave 2 model (called \textit{baseline}) with 8 NFE, the baseline with EDM methodologies (called \textit{EDM}) and FastWave model with acrchitecture changes and EDM methodologies (denoted as \textit{FastWave}).

The original NU-Wave 2 model, as confirmed by the official code repository of the paper, was trained for approximately 649 epochs using two NVIDIA A100s, representing a very high computational effort. We ran experiments in a limited mode, training each model for up to 30 hours on a single NVIDIA V100 GPU which represents more limited setting. To draw intermediate results in this limited mode, we also trained the NU-Wave 2 baseline for a full comparison. The intermediate comparisons are shown in Table~\ref{tab:intermidiate_models}. 


For the final comparison, we used the original NU-Wave 2 checkpoint from epoch 629 and also pushed our final version to epoch 140 on the NVIDIA 1xV100. 
For a more detailed comparison, we also validated FlowHigh \cite{flowhigh2025} and AudioSR \cite{audiosr} using the official implementations and our metrics, the results are presented in Table~\ref{tab:pretrained_models}. For training in all experiments we used Adam optimizer with learning rate 2e-4 with batch size 2.


\section{Main Results}

\begin{table*}[t]
\centering
\caption{Intermediate trained model comparison, the FLOPs are given for one function evaluation.}
\footnotesize
\label{tab:intermidiate_models}
\setlength{\tabcolsep}{4pt}
\begin{tabular}{|l|c|c|c|c|c|}
\hline
Metric
& Baseline 8 NFE
& EDM 4 NFE
& EDM 8 NFE
& FastWave 4 NFE
& FastWave 8 NFE \\
\hline
\multicolumn{6}{|l|}{\textbf{8 kHz}} \\
\hline
SNR $\uparrow$ & $17.47 \pm 4.33$ & $16.72 \pm 4.73$ & $16.02 \pm 4.60$ & $\mathbf{18.49 \pm 4.62}$ & $18.10 \pm 4.31$ \\
LSD $\downarrow$ & $1.31 \pm 0.12$ & $1.25 \pm 0.13$ & $\mathbf{1.21 \pm 0.10}$ & $1.22 \pm 0.12$ & $1.26 \pm 0.10$ \\
LSD-LF $\downarrow$ & $0.41 \pm 0.09$ & $0.39 \pm 0.09$ & $\mathbf{0.32 \pm 0.06}$ & $0.40 \pm 0.07$ & $0.36 \pm 0.06$ \\
LSD-HF $\downarrow$ & $1.42 \pm 0.13$ & $1.35 \pm 0.14$ & $1.32 \pm 0.11$ & $\mathbf{1.31 \pm 0.13}$ & $1.37 \pm 0.11$ \\
\hline
\multicolumn{6}{|l|}{\textbf{12 kHz}} \\
\hline
SNR $\uparrow$ & $20.20 \pm 3.90$ & $19.25 \pm 4.83$ & $18.38 \pm 4.88$ & $\mathbf{20.76 \pm 5.37}$ & $20.37 \pm 5.07$ \\
LSD $\downarrow$ & $1.22 \pm 0.11$ & $1.17 \pm 0.12$ & $\mathbf{1.09 \pm 0.08}$ & $1.14 \pm 0.11$ & $1.16 \pm 0.09$ \\
LSD-LF $\downarrow$ & $0.54 \pm 0.09$ & $0.53 \pm 0.10$ & $\mathbf{0.42 \pm 0.07}$ & $0.54 \pm 0.10$ & $0.46 \pm 0.07$ \\
LSD-HF $\downarrow$ & $1.36 \pm 0.13$ & $1.30 \pm 0.13$ & $\mathbf{1.23 \pm 0.09}$ & $1.26 \pm 0.13$ & $1.30 \pm 0.11$ \\
\hline
\multicolumn{6}{|l|}{\textbf{16 kHz}} \\
\hline
SNR $\uparrow$ & $22.26 \pm 3.55$ & $21.15 \pm 4.52$ & $20.22 \pm 4.84$ & $\mathbf{22.60 \pm 5.44}$ & $22.32 \pm 5.28$ \\
LSD $\downarrow$ & $1.14 \pm 0.11$ & $1.11 \pm 0.12$ & $\mathbf{1.01 \pm 0.07}$ & $1.07 \pm 0.10$ & $1.05 \pm 0.08$ \\
LSD-LF $\downarrow$ & $0.62 \pm 0.12$ & $0.65 \pm 0.13$ & $\mathbf{0.49 \pm 0.10}$ & $0.63 \pm 0.12$ & $0.51 \pm 0.09$ \\
LSD-HF $\downarrow$ & $1.30 \pm 0.12$ & $1.26 \pm 0.13$ & $\mathbf{1.17 \pm 0.08}$ & $1.21 \pm 0.12$ & $1.22 \pm 0.10$ \\
\hline
\multicolumn{6}{|l|}{\textbf{24 kHz}} \\
\hline
SNR $\uparrow$ & $24.43 \pm 3.24$ & $24.86 \pm 4.13$ & $23.69 \pm 5.29$ & $\mathbf{26.33 \pm 4.35}$ & $26.26 \pm 4.43$ \\
LSD $\downarrow$ & $1.01 \pm 0.11$ & $1.01 \pm 0.12$ & $\mathbf{0.86 \pm 0.06}$ & $0.95 \pm 0.08$ & $0.89 \pm 0.06$ \\
LSD-LF $\downarrow$ & $0.68 \pm 0.12$ & $0.73 \pm 0.14$ & $0.54 \pm 0.11$ & $0.70 \pm 0.12$ & $\mathbf{0.53 \pm 0.09}$ \\
LSD-HF $\downarrow$ & $1.21 \pm 0.12$ & $1.18 \pm 0.15$ & $\mathbf{1.09 \pm 0.08}$ & $\mathbf{1.09 \pm 0.10}$ & $1.10 \pm 0.09$ \\
\hline
\multicolumn{6}{|l|}{\textbf{Complexity}} \\
\hline
RTF $\downarrow$ & $0.26\pm0.02$ & $\mathbf{0.13\pm0.02}$ & $0.26\pm0.02$ & $0.15\pm0.10$ & $0.30\pm0.10$ \\
GFLOPS $\downarrow$ & $18.99$ & $18.99$ & $18.99$ & $\mathbf{12.87}$ & $\mathbf{12.87}$ \\
\#params $\downarrow$ & $1.8$M & $1.8$M & $1.8$M & $\mathbf{1.3}$M & $\mathbf{1.3}$M \\
\hline
\end{tabular}
\end{table*}

\begin{table*}[!ht]
\centering
\caption{Pretrained / large-capacity model comparison, the FLOPs are given for one function evaluation.}
\label{tab:pretrained_models}
\footnotesize
\setlength{\tabcolsep}{4pt}
\begin{tabular}{|l|c|c|c|c|c|}
\hline
Metric
& FastWave 4 NFE
& FastWave 8 NFE
& NU-Wave 2 8 NFE
& FlowHigh
& AudioSR \\
\hline
\multicolumn{6}{|l|}{\textbf{8 kHz}} \\
\hline
SNR $\uparrow$ 
& $\mathbf{18.75 \pm 4.84}$ 
& $18.53 \pm 4.73$ 
& $18.43 \pm 4.92$ 
& $18.04 \pm 4.74$ 
& $13.75 \pm 3.83$ \\

LSD $\downarrow$ 
& $1.18 \pm 0.12$ 
& $1.19 \pm 0.11$ 
& $1.15 \pm 0.10$ 
& $\mathbf{0.96 \pm 0.08}$ 
& $1.55 \pm 0.15$ \\

LSD-LF $\downarrow$ 
& $0.36 \pm 0.08$ 
& $0.28 \pm 0.05$ 
& $0.22 \pm 0.07$ 
& $\mathbf{0.24 \pm 0.02}$ 
& $0.44 \pm 0.07$ \\

LSD-HF $\downarrow$ 
& $1.27 \pm 0.13$ 
& $1.29 \pm 0.12$ 
& $1.25 \pm 0.11$ 
& $\mathbf{1.05 \pm 0.09}$ 
& $1.69 \pm 0.17$ \\
\hline

\multicolumn{6}{|l|}{\textbf{12 kHz}} \\
\hline
SNR $\uparrow$ 
& $21.08 \pm 5.71$ 
& $20.93 \pm 5.80$ 
& $20.95 \pm 5.18$ 
& $\mathbf{21.17 \pm 5.39}$ 
& $16.18 \pm 3.96$ \\

LSD $\downarrow$ 
& $1.09 \pm 0.11$ 
& $1.06 \pm 0.09$ 
& $1.02 \pm 0.08$ 
& $\mathbf{0.90 \pm 0.09}$ 
& $1.46 \pm 0.16$ \\

LSD-LF $\downarrow$ 
& $0.49 \pm 0.10$ 
& $0.38 \pm 0.06$ 
& $0.27 \pm 0.07$ 
& $\mathbf{0.28 \pm 0.05}$ 
& $0.55 \pm 0.13$ \\

LSD-HF $\downarrow$ 
& $1.21 \pm 0.13$ 
& $1.20 \pm 0.11$ 
& $1.16 \pm 0.09$ 
& $\mathbf{1.03 \pm 0.10}$ 
& $1.65 \pm 0.18$ \\
\hline

\multicolumn{6}{|l|}{\textbf{16 kHz}} \\
\hline
SNR $\uparrow$ 
& $23.07 \pm 5.85$ 
& $23.08 \pm 6.06$ 
& $23.31 \pm 5.17$ 
& $\mathbf{23.58 \pm 5.41}$ 
& $19.25 \pm 3.82$ \\

LSD $\downarrow$ 
& $1.04 \pm 0.10$ 
& $0.98 \pm 0.08$ 
& $0.94 \pm 0.08$ 
& $\mathbf{0.85 \pm 0.09}$ 
& $1.37 \pm 0.15$ \\

LSD-LF $\downarrow$ 
& $0.59 \pm 0.13$ 
& $0.44 \pm 0.08$ 
& $0.30 \pm 0.09$ 
& $\mathbf{0.28 \pm 0.05}$ 
& $0.54 \pm 0.13$ \\

LSD-HF $\downarrow$ 
& $1.17 \pm 0.12$ 
& $1.14 \pm 0.10$ 
& $1.12 \pm 0.09$ 
& $\mathbf{1.02 \pm 0.11}$ 
& $1.63 \pm 0.18$ \\
\hline

\multicolumn{6}{|l|}{\textbf{24 kHz}} \\
\hline
SNR $\uparrow$ 
& $27.09 \pm 4.84$ 
& $27.22 \pm 5.33$ 
& $27.68 \pm 4.21$ 
& $\mathbf{27.80 \pm 4.95}$ 
& $23.03 \pm 3.48$ \\

LSD $\downarrow$ 
& $0.93 \pm 0.08$ 
& $0.83 \pm 0.06$ 
& $0.78 \pm 0.06$ 
& $\mathbf{0.74 \pm 0.09}$ 
& $1.27 \pm 0.15$ \\

LSD-LF $\downarrow$ 
& $0.66 \pm 0.14$ 
& $0.48 \pm 0.09$ 
& $0.33 \pm 0.11$ 
& $\mathbf{0.30 \pm 0.06}$ 
& $0.58 \pm 0.15$ \\

LSD-HF $\downarrow$ 
& $1.08 \pm 0.10$ 
& $1.05 \pm 0.09$ 
& $1.04 \pm 0.08$ 
& $\mathbf{1.00 \pm 0.13}$ 
& $1.69 \pm 0.22$ \\
\hline

\multicolumn{6}{|l|}{\textbf{Complexity}} \\
\hline
RTF $\downarrow$ 
& $0.16\pm0.03$ 
& $0.30\pm0.14$ 
& $0.26\pm0.02$ 
& $\mathbf{0.06\pm0.02}$ 
& $4.99\pm1.59$ \\

GFLOPS $\downarrow$ 
& $\mathbf{12.87}$ 
& $\mathbf{12.87}$ 
& $18.99$ 
& $30.39$ 
& $2536.2$ \\

\#params $\downarrow$ 
& $\mathbf{1.3}$M 
& $\mathbf{1.3}$M 
& $1.8$M 
& $49.40$M 
& $1285.40$M \\
\hline
\end{tabular}
\end{table*}

\subsection{Limited Setting}

\noindent In this setting (see Table \ref{tab:intermidiate_models}) we observed that with the same training resource EDM methodology considerably improves convergence. We were able to outrun the training of \textit{Baseline} with \textit{EDM} in $30$ epochs reaching better reconstruction metrics in both 8 and 4 NFE settings. \textit{FastWave} demonstrated relatively the same performance as EDM reaching LSD below 1 in 24-48 task which is comparable to EDM. In terms of SNR FastWave models also show similar results  to NU-Wave 2 and FlowHigh, which indicates good phase reconstruction.

\subsection{Comparison with Baselines}

In Table \ref{tab:pretrained_models} we perform comparison with the recent diffusion baseline represented by Audio-SR (general checkpoint) and a flow-based model FlowHigh. FastWave and NU-Wave 2 considerably outperformed AudioSR  in all benchmark problems, this is most likely due to fundamental nature of Audio-SR where it was reported, that additional fine-tune is needed for speech. FlowHigh is the best in our comparison reaching LSD below 1 in all test problems. Despite being slightly worse than FlowHigh in LSD,  FastWave and NU-Wave 2 show better performance in SNR. It allows to place them in relatively the same performance class. FastWave has the smallest number of parameters among compared models and 2 times reduced NFE in comparison to NU-Wave 2. FastWave with 4 NFE it has only around 1.5 times more FLOPs than FlowHigh and much less than Audio-SR. It is important to note that our results with FastWave were achieved with significantly less training resource than all other considered approaches. RTF measurements demonstrate that FastWave has potential for streaming applications on consumer devices with GPU. 

\section{Conclusion}
In this work we presented an optimized diffusion-based pipeline for audio super-resolution from any to 48 kHz. Our approach requires moderate computational resources to train, it possesses small number of parameters and can be applied in low-resource setup, having the potential to work in  streaming mode.

\section{Acknowledgements}

This research was supported in part through computational resources of HPC facilities at HSE University.

\section{Generative AI Use Disclosure}
No Generative AI was used to prepare the text.

\bibliographystyle{IEEEtran}
\bibliography{mybib}

\end{document}